\documentstyle[aps,pre,psfig,12pt]{revtex}
\begin{document}
\textwidth 5.6in
\textheight 9.0in
\title{A simple and exactly solvable model for a semiflexible polymer chain 
interacting with a surface}
\author{P. K. Mishra, S. Kumar and Y. Singh }
 
\address{ Department of Physics, Banaras Hindu University, \\
     Varanasi 221 005, India }
\maketitle
\draft
\begin{abstract}
 We use the lattice model of directed walks to investigate the 
conformational as well as the adsorption properties of a semiflexible 
homopolymer chain immersed in a good solvent in two and three dimensions.
To account for the stiffness in the chain we have introduced energy barrier
for each bend in the walk and have calculated the persistent length as a
function of this energy. For the adsorption on an impenetrable surface 
perpendicular to the preferred direction of the walk we have solved the 
model exactly and have found the critical value of the surface attractions
for the adsorption in both two and three dimensions. We have also enumerated
all the possible walks on square and cubic lattices for the number of steps
$N\le 30$ for two-dimensions and $N\le 20$ for three dimensions and have used
ratio method for extrapolation. The transition located using this method
is in excellent agreement with the results found from the analytical method.
\end{abstract} 
PACS numbers: 64.60.-i,68.35.Rh,05.50.+q
\narrowtext
\section{Introduction} 
 
Biopolymers are known to exhibit under different environments a variety 
of persistent lengths ranging from
being much smaller than the over all length of the polymer, to being comparable
to the chain length $\cite{1}$. When the persistent length associated with the 
polymer is much smaller than the overall length of the chain, the polymer is 
said to be flexible. On the other hand, when the persistent length is comparable
to the chain length, the polymer is said to be rigid. When the persistent length
falls in between the two extremes, the chain is said to be semiflexible. 
The conformational properties of such chains have attracted considerable attention
is recent years because of experimental developments in which it has become 
possible to pull and stretch single molecule to measure elastic properties 
$\cite{2}$. Such studies reveal a wealth of informations about the conformational 
behaviour of semiflexible polymers that are of clear biological importance.

An impenetrable surface is known to affect the conformational properties of
polymers in a significant way $\cite{3,4}$. This is due to a subtle competition between
the gain of internal energy and a corresponding loss of entropy at the surface.
Since the flexibility of the chain affects this competition, a semiflexible 
chain is expected to show different adsorption behaviour compared to that of 
a flexible polymer chain. A stiff chain is known to get adsorbed easily compared 
to a flexible chain $\cite{5}$. 

A simple way to account the stiffness of a semiflexible chain is to constrain
the angle between the successive segments to be fixed. The value of the angle 
depends on the local stiffness of the chain. This prescription leads to the freely
rotating chain model $\cite{6}$. In the continuum limit the freely rotating chain
becomes the so called worm like chain $(WLC)$ $\cite{7}$. In these models the 
persistent length $l_p$ is defined as a characteristic length for tangent-tangent
correlation  function 
$<\underline{t}(s)\underline{t}(s^{'})>\simeq exp(-\frac{|s-s'|}{l_p})$. The tangent vector 
$\underline{t}(s)$ is defined as
$\frac {\partial \underline{r}(s)}{\partial t}$, where $\underline{r}(s)$ is parametrized in terms of the 
arc length $s$ of the chain $\cite{7}$. 

Though the worm like chain model of Kratky and Porod $\cite{7}$ has been 
used extensively
to study the conformational properties and surface adsorption of a 
semiflexible chain $\cite{8}$, it can not mimic exactly the dimensional behaviour of the real
chains. In this paper we use the lattice model of directed walk and introduce 
stiffness in the chain by associating energy with every bend of the walk and 
calculate the bulk and adsorption properties of the chain as a function of 
stiffness of the chain.

The paper is organized as follows: In Sec. II we describe the lattice model of
directed walk and investigate the bulk properties. We calculate the value of 
persistent length as a function of energy associated with the bend. In Sec. III
we discuss the surface adsorption of a semiflexible chain represented by a directed
walk on a plane perpendicular to the preferred direction of the 
walk. We also use the exact enumeration technique to locate the 
adsorption desorption transition and compare the result with those found
exactly. This allows us to comment upon the accuracy of the method of exact 
enumerations.

\section{A lattice model for the semiflexible chain}

We consider a model of a self-avoiding directed walk on a lattice $\cite{9}$. 
Though the directedness of a walk amounts to some degree of stiffness as all
directions of the space is not treated equally, stiffness in the chain is 
introduced by associating energy barrier with every turn of the walk. Though
the model is very restrictive in the sense that the bend can be either $90^{\circ}$ 
or no bend at all, The model can be solved 
analytically and therefore gives the exact values of conformational and 
adsorption properties of a semiflexible chain. We consider two specific 
cases of directed walks:
If the walker is allowed to take steps along $\pm y$-axis (in two-dimensions) and
only along $+x$-axis the walk is said to be partially directed-self-avoiding walk
$(PDSAW)$. On the other hand, if the walker is allowed only along $+y$ and $+x$
directions then the walk is said to be fully directed-self-avoiding walk 
$(FDSAW)$. In the case of three dimensions ($3D$), a $PDSAW$ is one in which
walker is allowed along $\pm y$-direction but only along $+x$ and $+z$ 
directions while in the $FDSAW$ the walker is allowed to move only 
along $+x$, $+y$ and $+z$ directions.

A stiffness weight $k=exp(-\beta \epsilon_{b})$ where $\beta=(k_B T)^{-1}$ is 
inverse of the temperature and $\epsilon_b(>0)$ is the energy associated with 
each turn. For $k=1$ or $\epsilon_{b}=0$ the chain is said to be flexible and for 
$0<k<1$ or $0<\epsilon_{b} <\infty$ the chain is said to be 
semiflexible. When $\epsilon_{b}\to \infty$ or $k\to 0$, 
the chain becomes a rigid rod.

The partition function of such a chain can be written as 

\begin{equation}
Z(x,k)={\sum}^{N=\infty}_{N=0}\sum_{ all walks  of N steps} x^Nk^{N_b}
\end{equation}

Here $N_b$ is the total number of bends in a walk of $N$ steps
and $x$ is the step fugacity. 

\subsection{Conformational properties of the chain in Two-Dimensions} 

(i){\bf The case of $PDSAW$}:
        
          In $2D$, the generating function for $PDSAW$  has two components; 
one along the directedness (${\it i. e. } +x$-axis) and 
other perpendicular to it (${\it i. e. } \pm y$-axis) as shown in Fig.1.
The recursion relation for these two components of generating function are 
$\cite{9}$;

\begin{equation}
X=x+x(X+2kY)
\end{equation}
\begin{equation}
Y=x+x(kX+Y)
\end{equation}

{\hspace{5.8cm}\psfig{figure=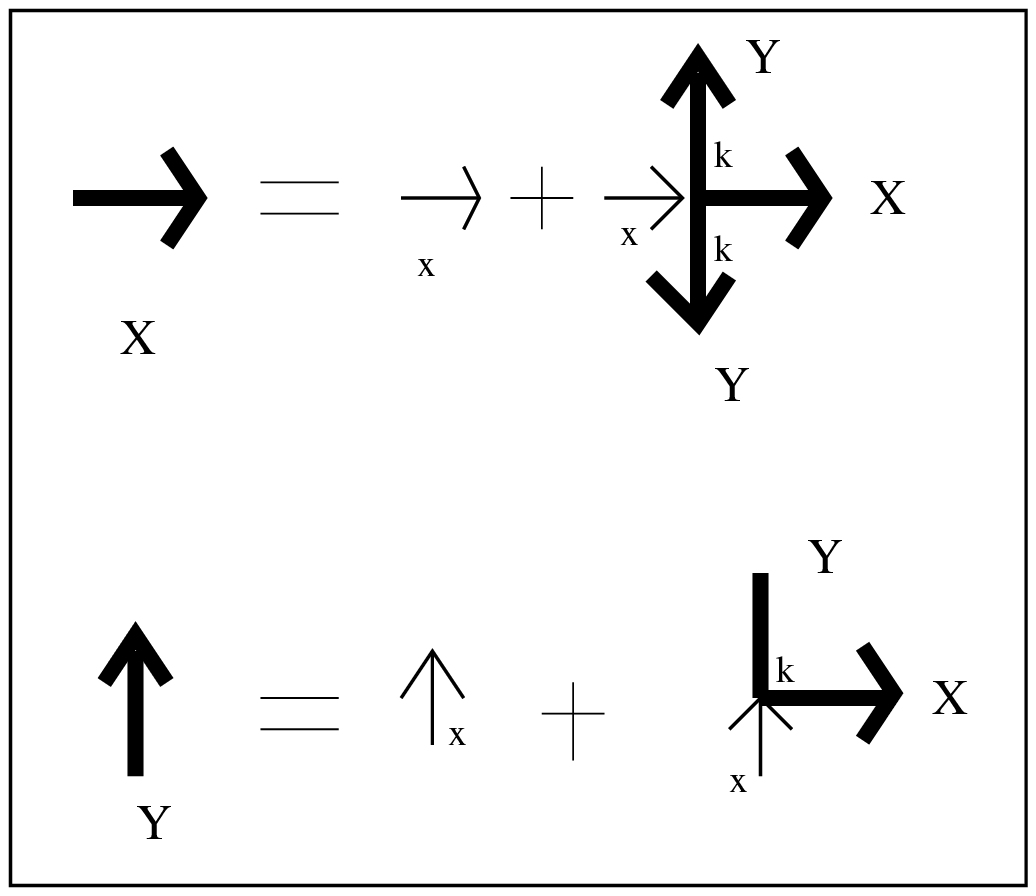,height=2.0in,width=2.0in}}

{\bf Fig. 1; \small The diagrammatic representation of the recursion relations 
Eqs. (2) and (3), for $PDSAW$. The thick arrows $X$ and $Y$
denote all possible walks with the initial step along $+x$ and $\pm y$ directions 
respectively.} 
\vspace{0.3cm}

Solving Eqs.(2) and (3) we get 

\begin{equation}
X=\frac{x+(2k-1)x^2}{1-2x+x^2-2x^2k^2}
\end{equation}
\begin{equation}
Y=\frac{x+(k-1)x^2}{1-2x+x^2-2x^2k^2}
\end{equation}

The partition function can therefore be written as 
\begin{equation}
Z^{2d}_{p. d.}(x,k)=X+2Y=\frac{(4k-3)x^2+3x}{1-2x+x^2-2x^2k^2}
\end{equation}

The critical point for polymerization of an infinite chain is found form
the relation 
\begin{equation}
1-2x+x^2-2x^2k^2=0
\end{equation}
This leads to the critical value of the step fugacity for a given value of 
$k$ as $x_c= \frac{1}{1+\sqrt 2 k}$. 

\vspace{0.5cm}
{\hspace{2.8cm}\psfig{figure=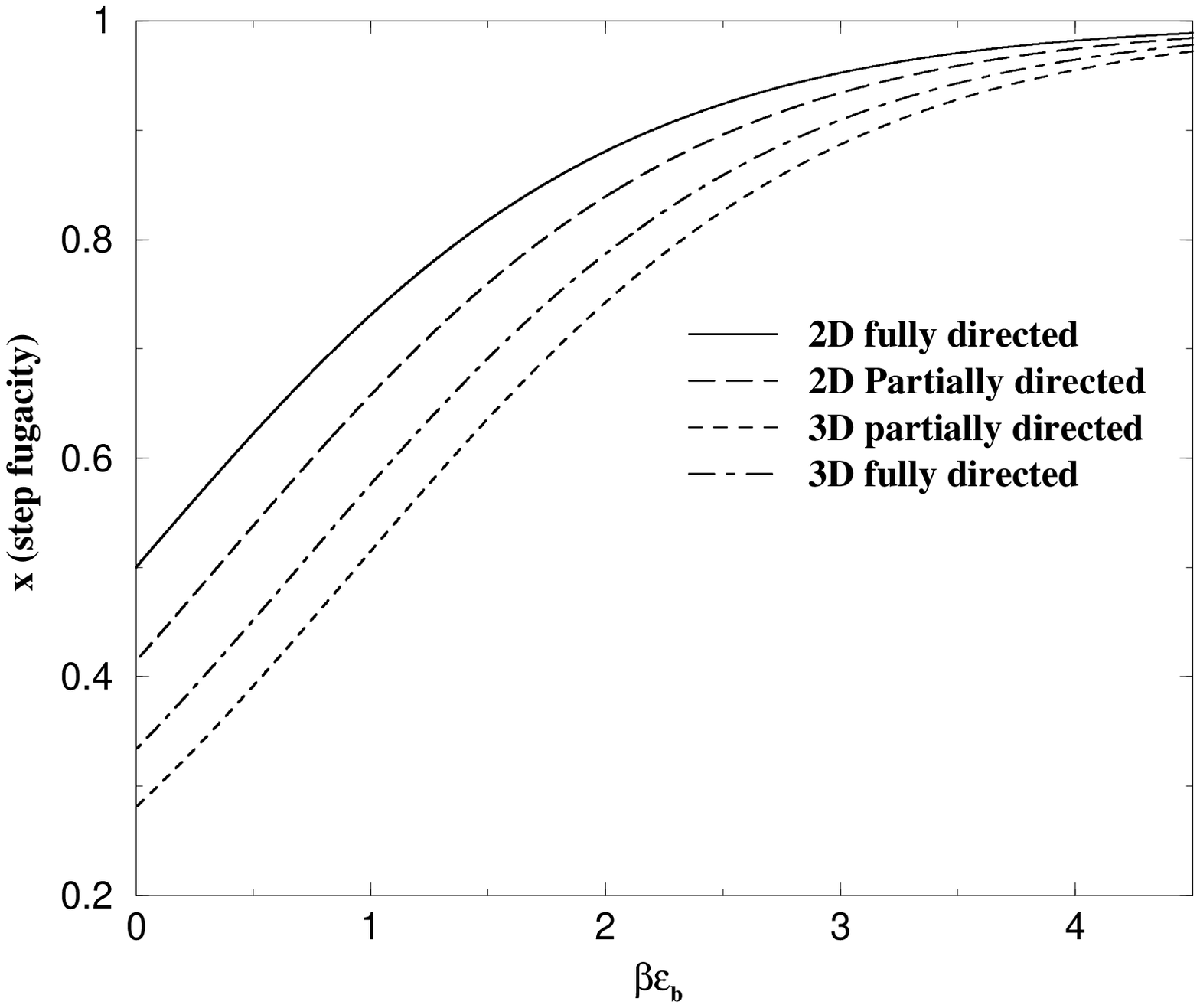,height=7.0cm,width=9cm}}

{\hspace{2.8cm}{\bf Fig. 2; \small The variation of step fugacity $x_c$ with 
$\beta \epsilon_b$.} 
\vspace{0.5cm}

The stiffness in the chain increases the
value of fugacity for polymerization. This dependence is shown in Fig. (2) 
by long-dashed line in which
we plot $x_c$ as a function of $\beta \epsilon_b$. We define the persistent length
as the average distance between two successive bends of the walk, ${\it i. e.}$

\begin{equation}
l_p=<L>/<N_b> 
\end {equation}

Where $L=<N>a$, $a$ being the lattice parameter.

For the $PDSAW$ in $2D$ we find

\begin{equation}
l_p= \frac{3+2\sqrt{2}}{4+3\sqrt{2}}[\sqrt{2}+exp(\beta \epsilon_b)]
\end{equation}

The dependency of $l_p$ on $\beta \epsilon_b$ is shown in Fig. (3) by 
a long-dashed line.

\vspace{1cm}
\hspace{3cm}\psfig{figure=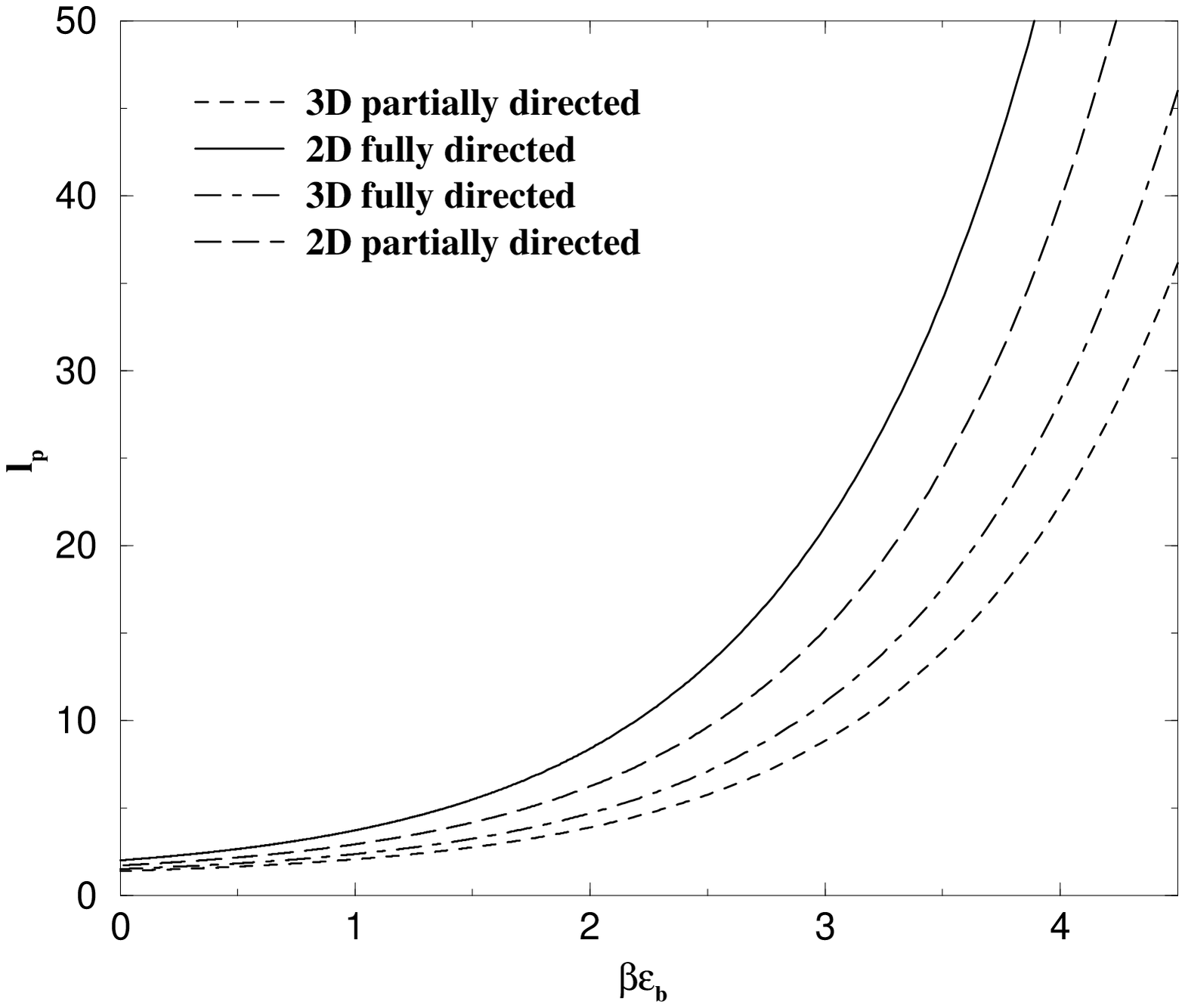,height=6cm,width=9cm}

\hspace{3cm}{\bf Fig. 3; \small The variation of $l_p$ with bending energy $\beta \epsilon_b$.}
\vspace{1cm}

The value of $l_p$ increases exponentially with the bend energy at a given 
temperature.

{\bf (ii) The case of $FDSAW$}:

In this case the polymer is directed along $+x$ and $+y$ direction; leading to
the following recursion relations for the generating functions,
 
\begin{equation}
X=x+x(X+kY)
\end{equation}
\begin{equation}
Y=x+x(kX+Y)
\end{equation}

Solving these equations we get the following value for the 
partition function $Z^{2d}_{f. d.}$;

\begin{equation}
Z^{2d}_{f. d.}=X+Y=\frac{2x}{1-(1+k)x}
\end{equation}

The critical value of step fugacity is found to be $x_c=\frac{1}{1+k}$.
The variation of $x_c$ with $\beta \epsilon_b$ is shown in Fig. (2) by solid
line.

The value of persistent length in this case attains a simple relation, 
${\it i. e.}$

\begin{equation}
l_p= 1+e^{\beta \epsilon_b}
\end{equation}
The variation of $l_p$ with the bending energy is shown in Fig. (3) by a solid 
line.

\subsection{Conformational properties of the chain in Three-Dimensions} 

(i){\bf The case of $PDSAW$}:

In the case of $PDSAW$ the polymer chain is directed in two-directions.
The recursion relations for generating functions are $\cite{9}$
\begin{equation}
X=x+x(X+2kY+kZ)
\end{equation}
\begin{equation}
Y=x+x(kX+Y+kZ)
\end{equation}
\begin{equation}
Z=x+x(kX+2kY+Z)
\end{equation}

Solving these equations we get the values of $X$, $Y$, $Z$ and the partition 
function as

\begin{equation}
X=Z=\frac{x+(2k-1)x^2}{(1+k-4k^2)x^2-(k+2)x+1}
\end{equation}
\begin{equation}
Y=\frac{x+(k-1)x^2}{(1+k-4k^2)x^2-(k+2)x+1}
\end{equation}
\begin{equation}
Z^{3d}_{p. d.}=X+2Y+Z=\frac{(6k-4)x^2+4x}{(1+k-4k^2)x^2-(k+2)x+1}
\end{equation}

That is at $x_c= \frac{k+2-\sqrt{17}k}{2(k+1-4k^2)}$, the $Z^{3d}_{p. d.}$ will
diverge. In this case the dependence of the fugacity for polymerization on the 
stiffness is more involved compared to the case in $2D$. The variation of $x_c$
with $\beta \epsilon_b$ is shown in Fig. (2) by the dashed line.
For persistent length we find 

\begin{equation}
l_p=\frac{[85+19\sqrt{17}-(102+26\sqrt{17})exp(\beta \epsilon_b)+(34+8\sqrt{17})
exp(2\beta\epsilon_b)]2[exp(2\beta \epsilon_b)+exp(\beta\epsilon_b)-4]}{(1-\sqrt{17}+2exp(\beta\epsilon_b))
[204+52\sqrt{17}-(272+64\sqrt{17})exp(\beta\epsilon_b)]+(85+21\sqrt{17})exp(2\beta \epsilon_b)} 
\end{equation}

The value of $l_p$ as a function of $\beta \epsilon_b$ is plotted in Fig. (3) by 
the dashed line.
\newpage
(ii){\bf The case of $FDSAW$}

In this case the polymer is directed along all the three directions {\it i. e.}
along $+x$, $+y$ and $+z$ directions. We can write following recursion relations
\begin{equation}
X=x+x(X+kY+kZ)
\end{equation}
\begin{equation}
Y=x+x(kX+Y+kZ)
\end{equation}
\begin{equation}
Z=x+x(kX+kY+Z)
\end{equation}
The solution of these equations leads to 
\begin{equation}
X=Y=Z=\frac{x}{1-(1+2k)x}
\end{equation}
Thus the partition function of the system can be
written as

\begin{equation}
Z^{3d}_{f. d.}=X+Y+Z=\frac{3x}{1-(1+2k)x}
\end{equation}
The critical value of the step fugacity is $x_c= \frac{1}{(1+2k)}$. The 
variation of step fugacity with bending energy $\beta \epsilon_b$
is shown in Fig. (2) by a dot-dashed line. In this
case $l_p$ is found to be 

\begin{equation}
l_p=1+\frac{1}{2}exp(\beta \epsilon_b)
\end{equation}
The value of $l_p$ as a function of $\beta \epsilon_b$ is plotted in Fig. (3) by 
a dot-dashed line.

In all the cases discussed above the persistent length shows exponential 
dependence on the bending energy.

\section{Surface adsorptions}

In the case of directed model we have two distinct surfaces; one parallel and the
other perpendicular to the directedness of the walk. The adsorption of polymer 
on a surface parallel to the preferred direction of the walk has been 
studied in case of $2D$ using the transfer matrix method $\cite{9}$. 
The features associated with the adsorption were
found to be same as in the isotropic case except that the critical value of surface
attraction for adsorption is higher. The surface perpendicular to the direction
of walk may give different features as the walk once leaves the surface it can not
return to it due to restriction on the walk. Here we report the result found 
analytically for the adsorption of directed semiflexible chain on a surface 
perpendicular to the directedness of the chain both in two and three dimensions.

\subsection{Adsorption of a directed semiflexible chain on a surface perpendicular 
to the directedness of the chain in two-dimensions} 

(i){\bf The case of $PDSAW$}:

In case of two-dimensions, surface is a line represented by $x=0$. Let $S$ be the
component of generating function along the surface and $X$ the component 
perpendicular to the surface as shown in Fig. (4). Following the method outlined
above we can write surface component as 

\begin{equation}
S=s+s(s+kX)+s^2(s+kX)+s^3(s+kX)+\dots
\end{equation}
where $s=\omega x$, and $\omega= exp(-\beta \epsilon_s)$ being the weight
associated with each step along the wall.

\vspace{0.3cm}

{\hspace{6cm}\psfig{figure=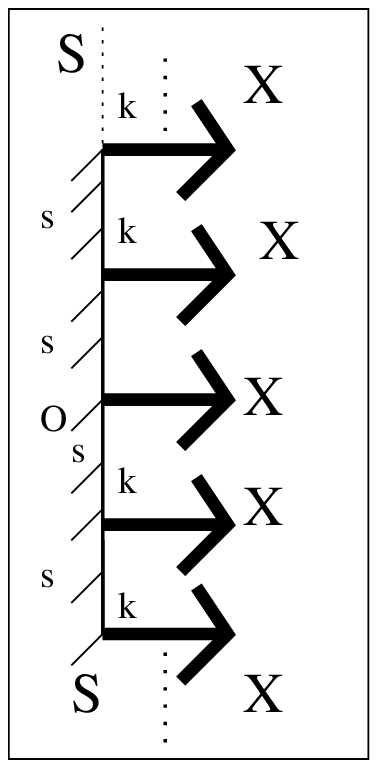,height=2.0in,width=1.0in}}

{\bf Fig. 4; \small The diagrammatic representation of the recursion relation 
(28). Each walk of the polymer chain starts from O. In 
this digram $X$ and $S$ denotes all possible walks with initial step
along the $+x$ and along the wall respectively.}
\vspace{0.3cm}

For $\omega=1$ Eq. (27) reduces to Eq. (5).

The partition function in presence of surface for $PDSAW$ is found to be

\begin{equation} 
{Z_s}^{2d}_{p. d.}(k,\omega,x) = X+2S
\end{equation}
Combining with Eq. (4) we find
\begin{equation}
{Z_s}^{2d}_{p. d.} =\frac{2sx^2(1-2k)+2s(1-2x)+(2sk+1-s)[x+(2k-1)x^2]}{(1-s)(1-2x+x^2-2k^2x^2)}
\end{equation}
The critical value of adsorption transition is found from the relation
\begin{equation}
(1-s)(1-2x+x^2-2k^2x^2)=0
\end{equation}
This leads to $\omega_c=\frac{1}{x_c}=\sqrt2 k+1$, which reduces to 
$\omega_c=\sqrt2+1$ $\cite{9}$ for the flexible polymer chain. 
The variation of $\omega_c$
with $\beta \epsilon_b$ is shown in Fig. (5) by a long dashed line.

\vspace{1cm}

\hspace{3cm}\psfig{figure=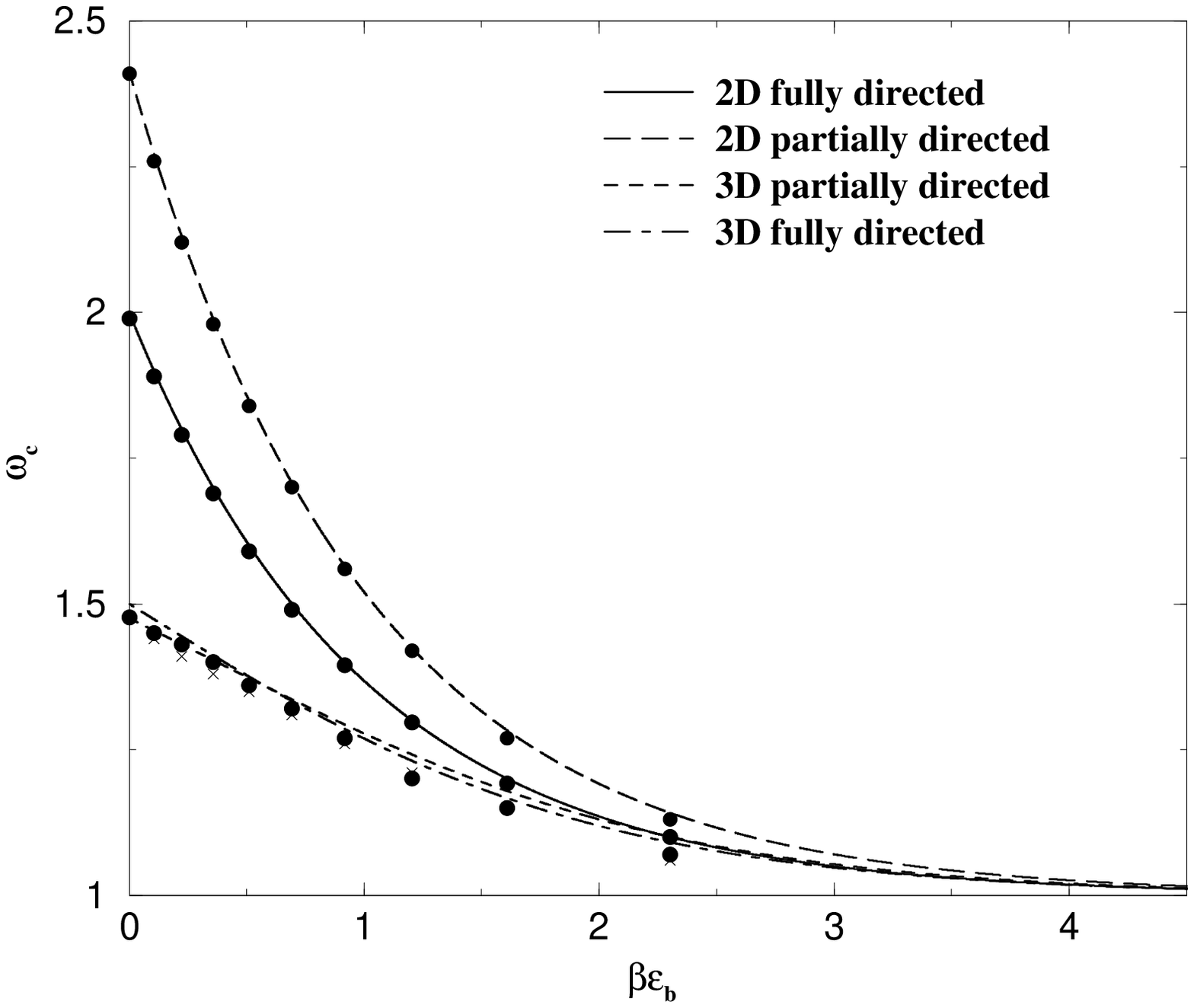,height=6cm,width=8cm}

{\bf Fig. 5; \small The exact value of $\omega_c$ for different values of 
$\beta \epsilon_b$. The lines in this figure correspond to analytical 
results however the dots on the lines correspond  to
the value obtained from exact-enumeration method in $2D$ and the cross used 
to denote the $\omega_c$ value for $3D$ partially directed case.} 

\vspace{1cm}

(ii){\bf The case of $FDSAW$}:

It is straight forward to show that for $FDSAW$ the partition function 
in the presence of surface is
\begin{equation}
S=s+s(s+kX)+s^2(s+kX)+s^3(s+kX)+\dots
\end{equation}
Using value of $X$ from Eq. (12) we get
\begin{equation}
Z_s^{2d}=X+S=\frac{s(1-(1+k)x)+x(sk-s+1)}{(1-s)(1-(1+k)x)}
\end{equation}

Which gives adsorption transition point at $\omega_c=1+k$. The variation of 
$\omega_c$ with $\beta \epsilon_b$ is shown in Fig. (5) by a solid line.

\subsection{Adsorption of a semiflexible directed chain on a surface perpendicular 
to one out of the two preferred direction of the chain in three dimensions}

(i){\bf The case of $PDSAW$}:

The analysis given above can be generalized in $3$ dimensions where surface 
dimension is two ${\it i. e.}$ $x-y$ plane at $z=0$. In the case of $PDSAW$
as mentioned above the choice of the walker is restricted to the $+x$-axis,
$\pm y$-axis and $+z$-axis.
Let $S_x$ and $S_y$ is the component of the
total partition function $Z_s^{3d}$ along $+x$ and $\pm y$ axis respectively,
however component perpendicular to the wall along $+z$ axis remains same as 
defined by Eq. (17).
We can, therefore write
\begin{equation}
S_x=\frac{s-s^2+2s^2k+Z(2k^2s^2+sk-s^2k)}{1-2s+s^2-2s^2k^2}     \hspace{3cm} (s<1)
\end{equation}
\begin{equation}
S_y=\frac{s-s^2+s^2k+Z(k^2s^2+sk-s^2k)}{1-2s+s^2-2s^2k^2}
\end{equation}
The expression for the partition function in this case found from the relation
\begin{equation}
{Z_s}^{3d}_{p. d.} = S_x+2S_y+Z
\end{equation}

Substituting the value of $S_x ,S_y$ and $Z$  we find
\begin{equation}
{Z_s}^{3d}_{p. d.}(k,\omega,x) =\frac{x(1-x+2kx)U+s(3-3s+4sk)V
}{(1-2s+s^2-2s^2k^2)[(1+k-4k^2)x^2-(k+2)x+1]}
\end{equation}
Where U and V are,
\begin{equation}
U =1-2s+3sk+s^2-3s^2k+2s^2k^2
\end{equation}
\begin{equation}
V=1-2x-kx+x^2+kx^2-4k^2x^2
\end{equation}

The two singularities appearing in eqn.($37$) give the critical value of 
$x_c = \frac{k+2-\sqrt{17}k}{2(1+k-4k^2}$ and 
$\omega_c = \frac{2(1+k-4k^2)}{(1+\sqrt2 k)(k+2-\sqrt{17} k)}$.
For flexible polymer chain (${\it i. e.}$ $k=1$ or $\epsilon_b =0$) 
it gives $x_c = \frac{-3+\sqrt{17}}{4}$
and $\omega_c = 1.47524....$.

(ii){\bf The case of $FDSAW$}:

For $FDSAW$, the partition function can be easily be evaluated. Here we write
the final form of the partition function as 
\begin{equation}
{Z_s}^{3d}_{f. d.}(k,\omega,x) = S_x+S_y+Z=\frac{(2s+x-3sx-3sxk)}{(1-s-sk)(1-(2k+1)x)} 
\end{equation}

The  two singularities appearing in Eq. ($39$) 
gives the critical value of $x_c = \frac{1}{2k+1}$ and $\omega_c =\frac{2k+1}{k+1}$.

The variation of $\omega_c$ with $\beta \epsilon_b$ are given in Fig. (5) 
for $PDSAW$ and $FDSAW$ by a dashed and dot-dashed lines respectively.

\section{Result from exact enumeration method}

Since the analytical approach is limited to very few cases, one often has to
resort to numerical methods, such as Monte Carlo simulations or a lattice
model using extrapolation of exact series expansion (referred to as exact
enumeration method). The later method has been found to give satisfactory
results as it takes into account the corrections to scaling. To achieve the
same accuracy by the Monte Carlo method, a chain of about two orders of 
magnitude larger than in the exact enumeration method has to be considered 
$\cite{10}$. 

We have enumerated all possible walks of length $N\le 30$ on square
lattice and of length $N\le 20$ on cubic lattice. The canonical partition 
function is written as
\begin{equation}
Z_N(k,\omega)=\sum_{N_s}\sum_{N_b} C_N(N_s,N_b) \omega^{N_s} k^{N_b}
\end{equation} 

Here $N_s$ is the number of steps on the surfaces. The reduced free energy
per monomer is found from the relation
\begin{equation}
G(k,\omega)=\lim_{N\to \infty} \frac{1}{N} log  Z_N (k,\omega)
\end{equation}
The limit $N\to \infty$ is found by using the ratio method $\cite{11}$ for
extrapolation.

The transition point for adsorption-desorption is found from the maximum of
$\frac {{\partial}^2 G(k, \omega)} {\partial{\epsilon_s}^2}(= \frac{\partial <N_s>}{\partial \epsilon_s})$. 
The transition points
found from this method are shown in Fig. ($5$) by dots and cross.
The results found from this method are in very good agreement with those 
found exactly 
in above sections. This result indicates that as for as locating the 
adsorption-desorption transition of a long flexible as well as semiflexible
chains immersed in a good solvents are concerned the method of exact 
enumeration can give reliable results.
\section {Conclusion}
In spite of the sever restriction imposed on the angle of bending of the
chain, the lattice models may provide interesting results for the conformational
and surface adsorption properties of a semiflexible chain. Introducing 
directedness in the walk allowed us to solve the model exactly in both two and
three dimensions. We have calculated the step fugacity for polymerization of
an infinite chain and the persistent length as a function of bending energy
associated with bending. We have also been able to obtain to the critical value
for adsorption of a directed chain on a surface perpendicular to the preferred
direction of the walk analytically in both two and three dimensions. The 
dependence of this critical value of surface attraction on the stiffness of 
the chain have been evaluated. 

We have also examined the accuracy of the method of exact enumeration in
locating the adsorption-desorption transition and have found that the method
give values that are in excellent agreement with the exact values.  

\newpage
\begin{center}
{\bf Acknowledgements}
\end{center}
This work has been financially supported by the Department of Science and 
Technology, New Delhi, $INDIA$.

\end{document}